# Policies and Economics of Digital Multimedia Transmission

Mohsen Gerami
The Faculty of Applied Science of Post and Communications
Danesh Blv, Jenah Ave, Azadi Sqr, Tehran, Iran.
Postal code: 1391637111

**Abstract**
There are different Standards of digital multimedia transmission, for example DVB in Europe and ISDB in Japan and DMB in Korea, with different delivery system (example MPEG-2, MPEG-4).This paper describe an overview of Digital Multimedia Transmission (DMT) technologies. The economic aspects of digital content & software solution industry as a strategic key in the future will be discussed. The study then focuses on some important policy and technology issues, such S-DMB, T-DMB, Digital Video Broadcasting– Handheld (DVB-H) and concludes DMT policies for convergence of telecommunications and broadcasting.

## 1. Convergence- the new pattern

Convergence is the key today. We can think about convergence in several different ways. One is in terms of the actual industries converging, such as communication, entertainment, and computing. Another is converging voice, video, and data over a common infrastructure or within a common computing platform.

One important factor during convergence is the transformation of the desktop computer through faster processors supporting advanced graphics and multimedia capabilities. The PC today is a collaborative communication and media tool.

Another factor driving convergence is the cost of maintaining three separate networks for voice, video, and data. Corporations can realize substantial saving in equipment, staff, and services by using converged networks.

Enterprises are looking at cost savings in the WAN as the first leverage point for convergence. Using Voice-over-Frame Relay, VOIP and Voice-over-ATM the same WAN lines can be used for voice as well as data, resulting in substantial cost savings.

Next, enterprises want to install the LAN and WAN infrastructure to do real-time video and audio information delivery.

Payload convergence is that aspect of converged networking wherein different data types are carried in the same communication format. However, the payload convergence does not prohibit the network form handling packets, according to their service requirement.

Protocol convergence is the movement away from multi-protocol to single protocol (typically IP) networks. While legacy networks are designed to handle many protocol and one type of data, converged networks are designed to handle one protocol and provided the services necessary for multiple types of data (such as voice, one way video, and interactive video).

Physical convergence occurs when payloads travel over the same physical network equipment regardless of their service requirements. Both multimedia and Web traffic can use the facilities of an edge network, even though the former has more stringent bandwidth, delay and jitter requirements than the later.

Device convergence means the trend in network device architecture to support different networking paradigms in single system.

Application convergence represents the appearance of applications that integrate formerly separate functions. For example, Web browsers allow the incorporation of plug-in applications that allow web pages to carry multimedia content such as audio, video, high-resolution graphics, virtual reality graphics and interactive voices.

Technology convergence signifies the move towards common networking technologies that satisfy both LAN and WAN requirements. For example, ATM can be used to provide both LAN and WAN services.

At the technical level, digital transmission has the potential to deliver integrated interactive text, video, voice and data to a mass audience what we might call 'real multimedia'. However, historically each part of the spectrum and mode of transmission became associated with a different form of communication: point-to-point communication became the province of the telephone, and wireless transmission became associated with broadcast news and entertainment. This is now changing with digital compression techniques the limited transmission capacity of existing telephone and cable infrastructures can be expanded to deliver a range of multimedia services that previously could only be carried on expensive broadband networks. Similarly digital compression allows point-to point communication to be conducted increasingly by advanced forms of radio transmission.

Telecommunications, media and IT companies are using the flexibility of digital technologies to offer services outside their traditional business sectors, increasingly on an international or global scale.

Convergence is not just about technology. Convergence is a debate about the impact of technology and a quantum leap towards a mature Information Society.

The Changes will offer many new opportunities for citizens to enrich their lives, not just the economic dimension, but the social and cultural ones as well. The global nature and interactivity of new communications media like the Internet are already opening new vistas, far beyond traditional, national media. Convergence will certainly expand the overall information market and be the catalyst for the next stage in the integration of the world economy. Even small business can market globally, thanks to the low cost of the World Wide Web site [1].

Innovations in technology, services and use are being driven by digitisation, higher-speed broadband networks and diversity in physical infrastructure, distributed connectivity and the emerging social web.

Developments in the diversity of physical infrastructure and broadband speed signal more choice, variety and increasing bandwidth. There are multiple distribution channels for professional content—mobile, IP-based, terrestrial and satellite broadcasting. Network access arrangements are likely to be a mix of open or shared access and closed systems. Advances in smart radio design and distributed connectivity is increasing the prevalence of wireless relative to wired access.

Inclusive of computer networking and IP-multimedia services, this trend is perhaps most notable for the integration of information-processing beyond the desktop into everyday objects and activities, or what is sometimes described as ubiquitous computing.

Key issues in this theme are the increasing use of content monitoring technologies, and the need to improve e-security and identity management. More recently, there has been a growing awareness of the potential improvements to energy efficiency and use from distributed micro-generation in 'smart-grids'.

Web-based technologies, falling under the umbrella term 'Web 2.0', have ushered in the era of user-generated content (UGC), and the continuous and seamless update of data and reusable services. For some years now, the web has enabled highly innovative developments at the edge. The web is now turning into both a platform and a database. Recent examples of this include the third-party development of applications programming interface (APIs), enabling interaction between programs on networked computers, mashups (content drawn from multiple websites) and widgets (portable code that any user can install and execute on their web pages).

In combination, these innovations are driving advances in computing processing power, display technologies, artificial intelligence and nanotechnology [2].

**The marketplace**

The technological revolution is widening access to communications links, permitting the distribution of commercial and entertainment material, and advancing telephony along the same highways. Digital technologies allow for a combination of different types of information representation, such as text, audio, images, and video, thereby removing the distinctions between different types of information production and distribution. Further, a variety of digital platforms will provide a range of services to users. Therefore, emerging information and communications networks may allow users to enjoy a virtually unlimited access to information, education, and entertainment resources, and reform the means through which people communicate to one another.

This convergence of information and communications systems (telecommunications, broadcasting, computers) has led to new opportunities for investment and innovation as the take up of information services within the business and the domestic sphere has grown. Thus, the production, distribution, and use of information has become an important economic activity as information's value as a commodity, which can be exchanged between service providers and consumers, has grown [3].

Convergence, in a descriptive sense, is often understood to take place at various levels. At a market level, convergence is associated with the increasing mergers, acquisitions and strategic alliances amongst corporations in broadcasting, telecommunications and IT. Industry lines seem to blur as various communications company seek to explore investments in non-traditional sectors. At this level, convergence is thought to raise economic issues of competition when increasingly oligopolies seem to emerge out the various joint ventures between the three communication sectors. There are also enormous challenges at the policy level whereby policies must be crafted to suit new competition issues raised by convergence such as threats of oligopolies as a result of merger activities. Interconnection pricing policy must also take into cognisance the converging networks [4].

## 2. Introduction to Digital Multimedia Transmission (DMT)

Digital transmission of multimedia has gained increasing importance in recent years.

Analogue TV transmissions are being replaced by digital multimedia transmission techniques, and mobile communication networks are evolving from pure telephony systems to third-generation mobile multimedia networks.

Nowadays it is more important combination of multimedia broadcast systems such as DAB and DVB with mobile multimedia networks such as UMTS.

Multimedia Broadcast/Multicast Service (MBMS) is a unidirectional point-to-multipoint service in which data is transmitted from a single source entity to a group of users in a specific area. The MBMS has two modes: Broadcast mode and Multicast mode.

MBMS broadcast mode: The broadcast mode is a unidirectional point-to-multipoint transmission of multimedia data (e.g. text, audio, picture, video) from a single source entity to all users in a broadcast area or areas. The broadcast mode is intended to efficiently use radio/network resources e.g. data is transmitted over a common radio channel. Data is transmitted to broadcast areas as defined by the network (Home environment).

MBMS multicast mode: The multicast mode allows the unidirectional point-to-multipoint transmission of multimedia data (e.g. text, audio, picture, video) from a single source point to a multicast group in a multicast area. The multicast mode is intended to efficiently use radio/network resources e.g. data is transmitted over a common radio channel. Data is transmitted to multicast areas as defined by the network (Home environment). In the multicast mode there is the possibility for the network to selectively transmit to cells within the multicast area which contain members of a multicast group [5].

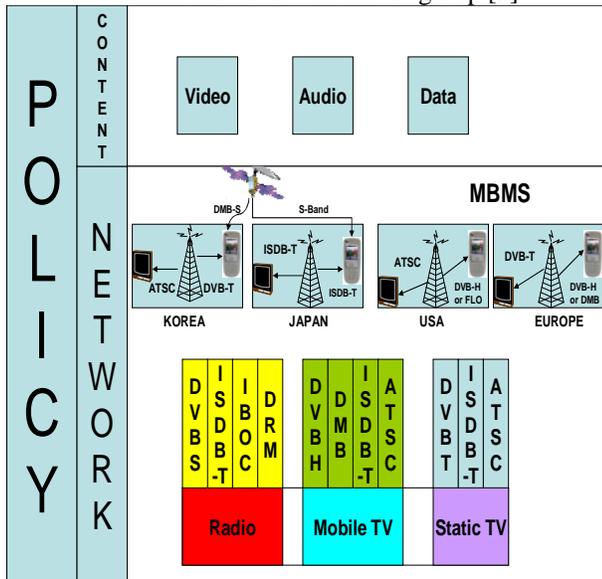

Figure 1.Digital Multimedia Transmission

The Eureka-147 DAB is decided as an official transmission system of Korean DMB mainly targeting the mobile TV service [6].

DMB is seen as the next generation digital broadcasting service for indoor and outdoor users. The DMB users currently can enjoy CD quality stereo audio services and real-time video/data streaming services anywhere in the nation while moving at the speed of up to 200 km/h (ITU). The audio service in DMB should support the standardized stereo audio broadcasting at the sampling rates of 24, 44.1, or 48 KHz. The service should provide CD quality audio for the audio-only broadcasting and better than the analog FM radio quality for the audio accompanying the video. The maximum bit-rate for the audio data in stereo is set to 128 Kbps [7].

Satellite Digital Multimedia Broadcast (S-DMB) is a mobile satellite system which allows for mass-marketed communication medium in different areas, where terrestrial coverage is not available. S-DMB is suitable in cases where terrestrial networks have not been deployed for business attractiveness reasons, or because terrestrial system has negative impacts on environment (crisis conditions), or where terrestrial coverage requires extensive investment programmes.

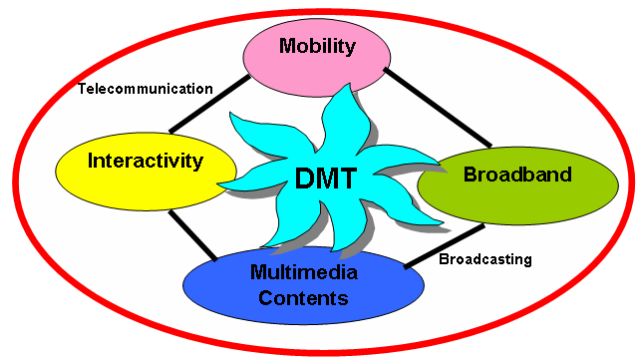

Figure 2.Digital Multimedia Statement

S-DMB provides a high data rate broadcast service to any 3GPP standardised class 3 handsets (250mW).The system operates in the IMT-2000 (International Mobile Telecommunication 2000) frequency band allocated to Mobile Satellite Services and is based on a dedicated high power geostationary satellite [8]. In this study we will further analyse multimedia transmission technologies, content, business and policies.

## 3. State-of-the-art broadcasting technology

This chapter contains technical description of different elements and components of the DMT technologies, their properties and functions. I focus on the major technologies of DMT.

## 3.1 DMB Networks

Subheadings should be as the above heading "2.1 Subheadings". They should start at the left-hand margin on a separate line. Digital multimedia broadcasting or DMB technology enables people on the go to enjoy crystal-clear video, theater-quality audio and data through handheld devices like handsets or in-car terminals. All these wonders are made possible through digital broadcasting technology. The conventional TV reception scheme, based on an analog technology, produces various noises and broken images with a mobile device. The DMB technology has created innovative solutions to problems. Technological challenges to local engineers were great, though. The most important factor in DMB service is speed. It requires an enormous amount of data to be processed quickly for seamless video and audio service. The speed is especially important for providing stabilized, quality service to portable terminals.

Digital multimedia broadcasting (DMB) uses the DAB mechanism to broadcast multimedia content to mobile receivers. The advent of H.264 and scaled down video requirements have made it feasible to use this robust transmission system for mobile environment meeting many of its exacting requirements [9].

Tailored content, adapted to mobile environments, high standards of sound/image quality and Security issues, preventing virus attacks to personal data are Multimedia acceptance by end- users.

Cost effective content delivery and nation wide coverage to maximize audience and Protection against unauthorized content sharing are major operators concerns.

Largely inspired by personal video recorder (PVR) service implementation, the S-DMB system depicted in Figure below provides anytime, anywhere service consumption to the end-users. The multimedia content is conveyed to the 3G handset local cache through a direct satellite distribution link exploiting the 3GPP push service over MBMS, whereas service management and interactivity is achieved through the 3G mobile network.

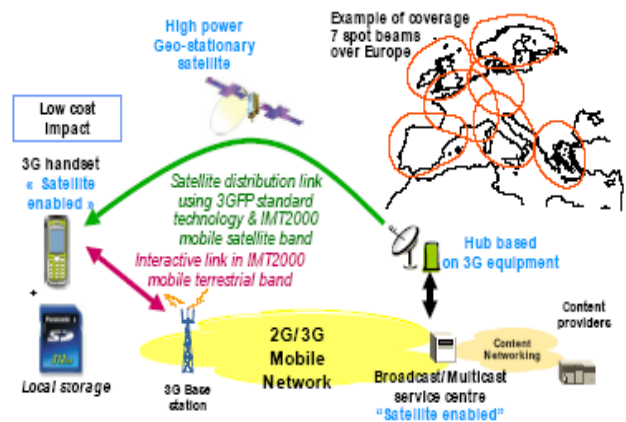

Figure 3.S-DMB concept architecture to enhance MBMS delivery on 3G and beyond 3G systems

Local storage combined with push service technology optimizes the bandwidth usage over the whole day duration and maximizes the satellite broadcast content delivery capacity: local storage can be filled at any time of day and regularly updated with predictive cache management techniques providing mobile operators with an increased content delivery capacity. In each handset, only those contents with a potential interest for the user are processed at physical level and stored to be available upon user's request. Streaming services are delivered using the 3GPP MBMS streaming by pre-empting the capacity allocated to push & store services [10].

## 3.2 T-DMB Technology

T-DMB has been introduced in Korea for digital radio services based on Eureka 147 called as DAB (Digital Audio Broadcasting) with an extension of multimedia services in the VHF band, named as Terrestrial Digital Multimedia Broadcasting (T-DMB).

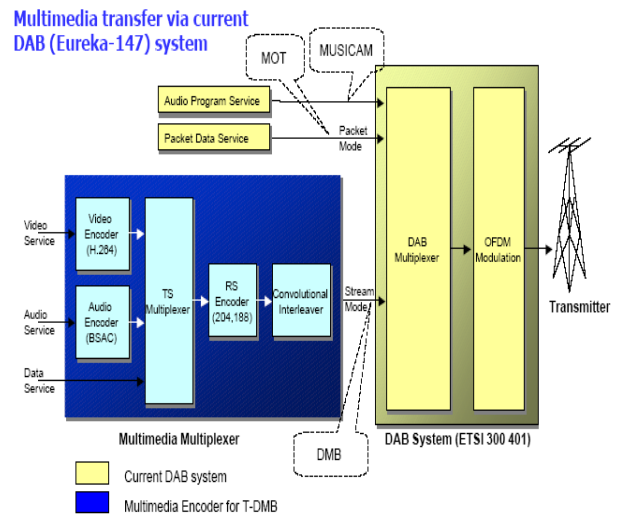

Figure 4.T-DMB Transmission system -source: [11].

T-DMB provides the traditional audio services through DAB framework and also multimedia ones with VCD visual and CD like quality. Beyond AV services, T-DMB can provide data services dependent or independent on AV programs. Also, those data services can be realized by a user's interaction through local and remote environments through telecommunication channels.

Korean T-DMB adopted MPEG-4 Advanced Video Coding (AVC) as a video compression standard to provide a high compression performance in the given bandwidth restriction.

Major components of T-DMB system are ITU-T H.264 | ISO/IEC 14496-10 MPEG-4 AVC (Advanced Video

Coding), for video, ISO/IEC 14496-3 MPEG-4 BSAC (Bit-Sliced Arithmetic Coding) for audio and ISO/IEC 14496-1 MPEG-4 System for interactive multimedia. ISO/IEC 13818-1 MPEG-2 TS is used for multiplexing of contents and a combination of convolutional interleaver and Reed-Solomon (RS) coder is used to improve BER performance.

The MPEG-4 AVC is known to have excellent compression efficiency as high as up to twice that of MPEG-4 Part 2 Visual (ISO/IEC 14496-2.) In the subjective test conducted by the working group, AVC demonstrated VCD-like quality at around 384 kbps.

The MPEG-4 BSAC is known to have the same compression efficiency as MPEG-4 AAC (Advanced Audio Coding) and characterized with additional functionality of fine grain scalability. In the subjective test conducted by the working group, BSAC demonstrated FM-like sound quality at 64kbps/stereo and CD sound quality at 96kbps/stereo.

In addition, MPEG-4 Systems was adopted to provide interactive services with appropriate synchronization among the multimedia objects. The Binary Format for Scene (BIFS), a part of MPEG-4 Systems, provides flexible composition capability for various types of multimedia objects in conjunction with MPEG-4 Synchronization Layer (SL) which enables a smooth rendering of different types of multimedia objects [12].

### 3.3 S-DMB Technology

Satellite DMB is a new concept in multimedia mobile broadcasting service that facilitates convergence between telecommunications and broadcasting.

The satellite DMB frequency is the most suitable for the mobile environment among those frequencies available for satellite use. As interesting property is that its frequency quality not affected by rainfall.

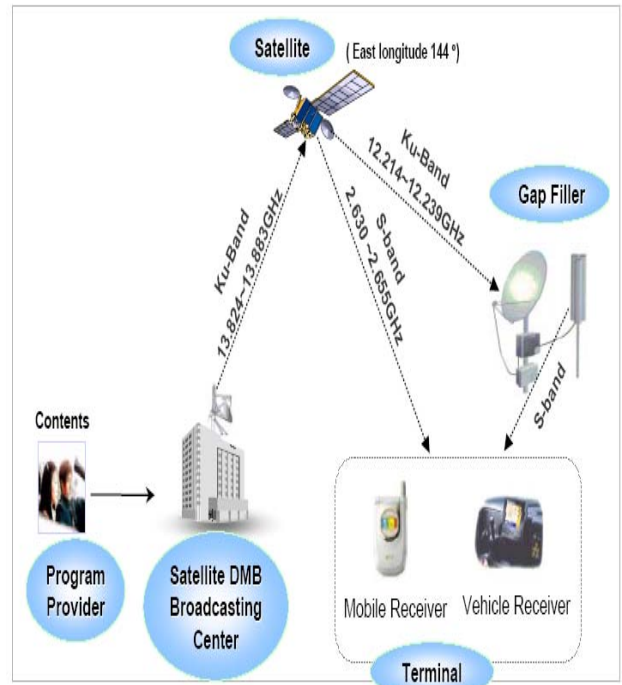

Figure 5.S-DMB network structure  (Source: SK telecom)

The satellite DMB service adopts the same Code Division Multiplexing (CDM) technology as the mobile phone service. Thus it is the most appropriate for signal reception in a mobile environment. This can also guard against multiple channel interferences that cause reductions in signal receiving quality within the mobile environment [13].

### 3.4 T-DMB vs. S-DMB

Based on the European 'Eureka-147' standard, the Korean T-DMB system's TV channels
7 through 13 utilize the 174-216MHz VHF band. Channel 12 has been allocated for DMB and is divided into three frequency blocks. Each block, or multiplexer, is allocated
1.54MHz of bandwidth that is capable of receiving one channel of video and three channels of audio or data.

The ITU-R BO 1130-4 'System E' is the standard for the Korean domestic S-DMB system. The original band allocation for domestic S-DMB (2.605-2.630GHz) recently was increased to the entire 50MHz range of 2.605-2.655GHz. It also has been determined that the T-DMB system can operate three channels of video and nine channels of audio/data, and that SKT can operate 39 channels of S-DMB; 11 video, 25 audio, and three data [14].

| | T-DMB | S-DMB |
|---|---|---|
| Technology transmission standard | * Eureka 147 (DAB-based) | * System E (similiar to CDMA) |
| Frequency | * CH12(204-210 Mhz) in Band III<br>* Considering additional CH8 and CH10 | * S-band (2.605~2.655 GHz) |
| Channels available | * 1 VHF TV channel can allocate 3 blocks and one block can cover 1 video and 3 data/audio channels | * About 13 video channels possible in 25MHz bandwidth<br><br>* For SKT, it plans to operate 11 video, 25 audio and 3 data channels |
| Mobile reception | * Available on the ground (in building and underground are not guaranteed)<br><br>* Need solution for gap areas | * Possible(planned to solve gap)<br><br>* S-band power control and gap fillers are needed for direct reception from ground |
| Display | * Not decided (7-15 inches considered) | * 7 inches maximum |
| Business profit model | * Free service through advertising revenue | * Paid service |
| Service coverage | * Local broadcasting | * Nationwide broadcasting |
| Target market | * Individuals or vehicles | * Individuals or vehicles |
| Cost | * About $43M | * $340M - $680M: depending on investment on gap fillers |

Source: In-Stat/MDR 8/04

Figure 6.features of T-DMB and S-DMB

## 3.5 IPDC

"IP data casting is a broadcast technology which enables cost effective and efficient distribution of digital content to mass audiences" [15].

IP data casting is a combination of digital broadcasting, IP based services and multimedia content. It enables a large scale distribution of multimedia content to users. The terminals are not limited to handheld devices. Therefore different opportunities can be explored such as in-car entertainment [16].

Delivery with IPDC has a fixed transmission cost. It costs the same amount of resources to broadcast content regardless of the number of users. On the other hand, the performance of transmission of multimedia content over 2.5/3G networks or Wireless Local Area Network (WLAN) hot-spots depends on the amount of users.

Broadcast transmissions reach all its audience without bottlenecks and without dependence on the number of users [17].

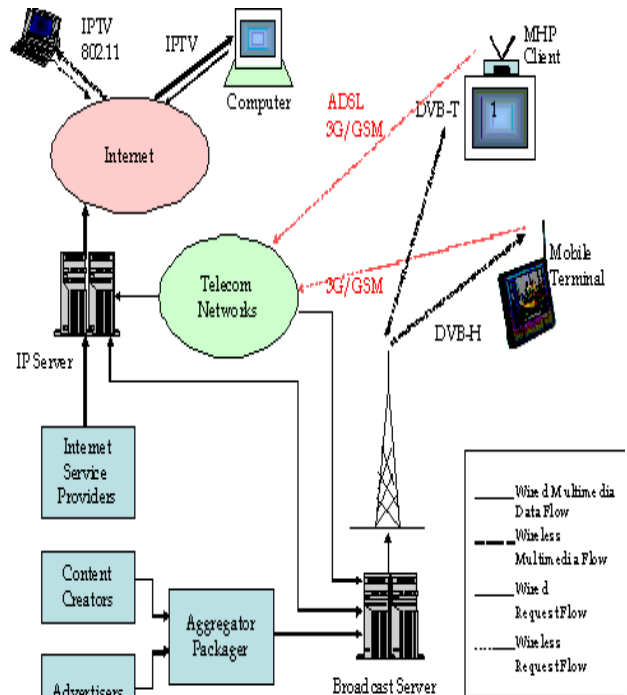

Figure 7.Interactive multimedia services over combined broadcasting and telecom networks

The main differences between IPDC aimed for mobile terminals and digital television are: built-in antennas, small screen and reliance on battery power. With the use of DVB-H technology, power savings of up to 90% can be achieved.

IPDC provides the capability to send 50-80 channels (128-384Kbps) over the same network vs. 3-4 channels (2-5Mbs) in digital television.

Due to the lack of battery power mobile terminals have, IPDC relies greatly on bandwidth transmission efficiency. The compression mechanisms selected have to be optimal yet provide flexibility to suite different terminals. The compression should reduce the file size as much as possible to minimize the transmission times, and the decoder should not be too complex. Otherwise, decoding the compressed file will consume more battery power.

IPDC value chain:

First, the content is created by the content producer. This content is delivered to the content provider who may modify or alter the content before selling it to the content aggregator. The content aggregator then schedules the content and creates combination of services that are delivered to the IPDC service operator.

The IPDC service operator encrypts the content, creates the access rights objects and develops an electronic service guide as well as price listings. The resulting stream is broadcasted in the air by the broadcast network operator. The user receives the stream and consumes the services.

The services may be accessible directly while some others have to be ordered through a different communication channel provided by the interaction channel service operator. Lastly, the interaction channel network operator maintains the physical interaction channel network, for example GSM or W-CDMA [18].

## 4. DMT Content

### 4.1 Digital Content

The digital content & software solution industry is a key strategic industry in the future that creates new demand and strengthens the competitive edge of other industries.
Core technologies such as 3D computer graphics, multi-platform geared online game engines and multi-platform e-learning solutions will be developed.
The digital content is comprised of text, image, sound that is digitized and serviced to subscribers. Representative digital content include games, videos, mobile and an e-learning content.
Digital content is unique in that it is possible to create value through 'One Source Multi Use (OSMU)'. Due to increased distribution of wired and wireless enabled handsets and the availability of application for these handsets, the value of mobile phone, PC, telematics, DTV and related industries can increase [19].

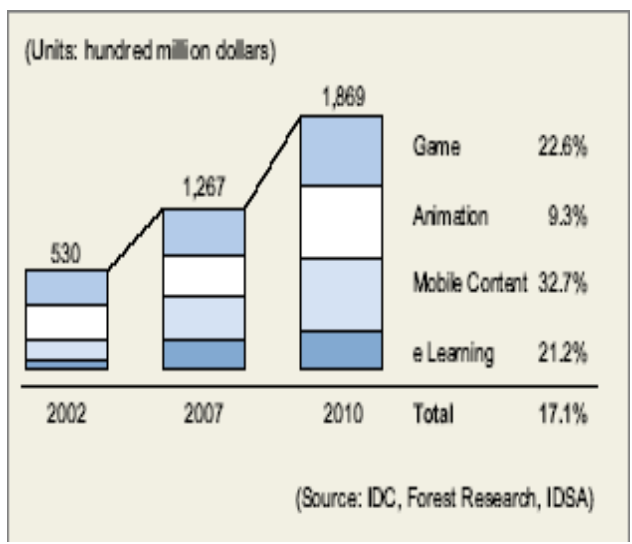

Figure 8. Worldwide Digital Market size

Content is the most important property of a broadcasting product. Several factors need to be considered while devising content for the mobile applications. Limited attention span of mobile consumer, size and clarity (resolution) of the display device and fidelity of the sound device, battery capacity of the device and lifestyle of mobile consumers as a group and their service requirements are important factors.
Creation of an entirely new genre of content is the only way forward, incorporating new concepts and approaches that appeal to audiences on the move and are suitable for the mobile devices they possess.
Re-purposing the existing or archived content is an excellent way of cutting down the costs. It involves re-dimensioning the content in accordance with the new parameters; short duration, smart editing, captivating audience interest and motivating them to seek more content are some such factors [9].
3G mobile television should prove to be the best method of viewing on-demand, streamed, or short-format video; broadcast mobile television would then establish itself as a genuine mobile television service aimed at a mass audience.
Moreover, mobile television has long been presented as the 3G "killer application", but it is just one of the many possible 3G services and applications – along with position determining technology (as a supplement to a GPS service), information consumption (news, audio and video), downloads (software, games and ringtones, etc.), messaging (instant or MMS, etc.), videoconferencing and internet access [20].

### 4.2 Economics of content/information

The mobile content industry is still in its formative stages, and its value chain and business models are unsettled and complex. Numerous players are vying to control parts of value chain. M-Commerce already has the technical capability to deliver services but requires cooperation of telcos, banks, and merchants. "Trigger" for the widespread adoption will come when all the major players and some major banks sign up to a common industry-wide platform (especially with mobile wallet). Mixing of financial and mobile value chain because difficulty of mobile provider to become a bank.
Revenue models and revenue split for mobile content vary widely and industry continues to experiment. Users typically pay data fees to use the network as well as a separate fee for purchase of content. Mobile operators need to increase the average revenue per users (ARPU), because the costs of content. One alternative is subscription basis [21].
Digital walkmans will be one of the challenges in the impending battles amongst industrialists. Will they, in logic of convergence and reduction, become more like mobile telephones? Some MP3 player manufacturers are joining forces with the mobile telephone industries in order to produce a hybrid terminal (e.g. Apple with Motorola). Or will they, in logic of divergence and diffraction, be turned into video mobile devices with huge memories and a video screen?

With increased bandwidth on both fixed and mobile networks as a backdrop, the PC-MP3 player combination can compete with the mobile television packages on a permanent basis. Mobile games console manufacturers face the same dilemma of convergence versus divergence.

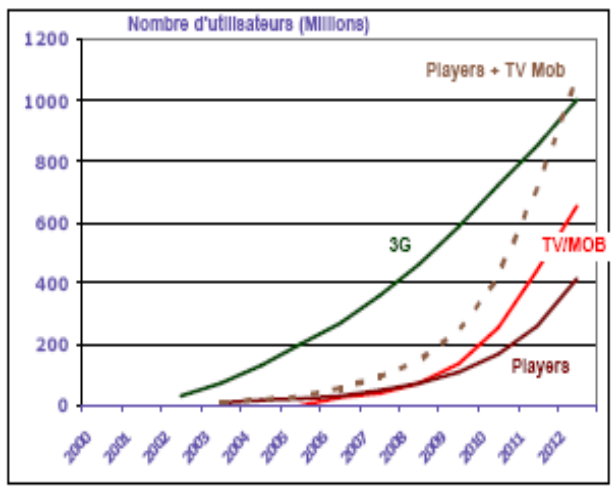

Figure 9.Growth of Digital Contents

Although they are sure about the technology's validity, operators do not necessarily have clear ideas about the economic model associated with this new service. The TV mobile battle will not only concern tools and pipes but also content.

Tests in Japan and Korea have not yet resulted in a viable economic model which will enable everyone (telecommunications operators and television package providers) to make a profit.

The challenge is significant for key-players in the content field. They are waiting for this new broadcasting method to guarantee them the level of income that they think they deserve in return. But for all that, the value chain, and particularly the distribution of value, will have to take into account what users will accept on their bill.

As a result, uncertainties about the economic model give rise to several very different sorts of questions:

Is mobile television a new media? Will it grow in harmony with the existing traditional television packages or will it be in competition with them?

What exactly do we know about the consumer expectations? Will they be prepared to pay for new services as an addition to an existing bill or as a completely new service?

What can we learn (use services, programmes and formats) from the first phase of the 3G rollout in Europe and the experience of operators in Japan and Korea which dates back further? Which programmes and services were viewed most?

How do users assess these new services?

Operators' initial packages have various coexisting billing methods: payments by session, by download, by volume and differentiation for peak and off-peak times.

These billing methods are not necessarily clear to consumers.

How do network operators and programme publishers negotiate their respective roles in the value chain?

Will providers of software solutions (particularly those for managing digital rights) and equipment manufacturers be able to gain a foothold in a value chain which is far from stable?

Will the new players who specialize in supplying mobile content, pioneers and innovators be able to hold their position in a game that seems to give priority to the "big boys"? [20]

It is generally known that there are different interconnected groups of ICT development factors. These include the issues of connectivity, access, capacity, regulation and socio-Cultural Environment. Developing a strong physical infrastructure and network connections is not sufficient unless this is accompanied by adopting national and international policies that aim to sustain universal, ubiquitous, and affordable access to ICT services. National and international strategies for ICT development should not ignore the socio-cultural aspects of diverse societies [22].

## 5. CONCLUSION

Convergence of broadcasting and telecommunications are in their early stages. Convergence basically describes two trends: the ability of different network platforms (broadcast, satellite, cable, telecommunications) to carry similar kinds of services; and the merging of consumer devices such as telephones, televisions, or PCs.

Today's platforms already support multi-play service line-ups that give operators a competitive advantage, and as platforms continue to evolve toward multi-service convergence, competitive advantages that they enable will increase [23].

In addition to the business drivers there are a number of other factors driving progress in the DMB technology market further. People are looking for a life that is more enriched and cultural, more flexible and diversified, more comfortable and safe, and more personal and convenient. Advancements in communication and information technology will be a major factor in realising these and customized needs.

A convergence of ubiquitous computing, ubiquitous communication and intelligent user interfaces will bring us towards the Ambient Intelligence era.

Digital convergence thus requires the convergence of public policies. On the regulatory level we face problems of a sectorial approach, which repeatedly regulates only a narrow field, lacking a broader understanding of how this field may relate to other fields. On the other hand, regulations need to be based upon numerous standards enforced on an international level (because only this makes them meaningful). In spite of their political declarations in support of the development of information society, most governments do not seem to have fully solved the emerging issues in the field of public policies which regulate the convergence of content, communication and information industries [24].

Broadcast technologies are changing fundamentally. Technology will get better, smaller, and cheaper by the day. This will spike off a new wave of innovation in content and build new business models with multiple platforms on offer. New distribution platforms will evolve the relationship between content and carriage. The lowering of costs and multiple platforms like digital cable, DTH, IPTV, and mobile TV will keep fueling more and more broadcast TV channels and niche content [23].

It can be said that we are moving towards a more natural use of network services in different branches of life. This view incorporates the advancements in technologies encouraging users to collaborate and share content, and use personalized and increasingly intelligent services. As the amount of available services and content increases, there will be functions for helping the user to manage services and content automatically based on his context or profiles.

It is necessary to work out convergence of services which covers the area between telecommunications and broadcasting and establish policy coordination mechanism to improve regulatory efficiency

It is important to facilitate Sharing of transmission lines and investigate the possibility and opportunity of separate regulation on networks (including service) and contents respectively.

It is required to improve law and current regulatory system that impede revival of converged services already available from technical development.

## 6. Appendix

**Advanced Television System Committee (ATSC)** is a standards organization based in the United States to promote the establishment of technical standards for all aspects of advanced television systems. The digital terrestrial television standard developed by ATSC specifies 8-VSB transmission for HDTV (www.atsc.org).

**Digital Multimedia Broadcast (DMB)** delivers mobile television services using the Eureka-147 Digital Audio Broadcast (DAB) standard with additional error-correction. T-DMB (terrestrial) uses the terrestrial network in Band III and/or Band L while S-DMB (satellite) uses the satellite network in Band L (World dab, www.dvb.org).

**Digital Video broadcasting standard for terrestrial television (DVB-T)** is a transmission specification for digital terrestrial television developed by the DVB Project. The **DVB Project** is an industry-led consortium of over 260 broadcasters, manufacturers, network operators, software developers, regulatory bodies and others in over 35 countries committed to designing global standards for the global delivery of digital television and data services (World dab, www.dvb.org).

**Eureka 147** is a standard for digital audio broadcasting (World dab, www.dvb.org).

**Integrated Services digital broadcasting (ISDB)**
In Japan, NHK has proposed an integrated broadcasting system, which can be used for terrestrial, satellite and cable applications. The development is known as Integrated Services Digital Broadcasting (ISDB). The system has been designed to carry video, sound and a wide range of data in one broadcasting channel and ISDB will use standards common to computer and telecommunications networks.

For the terrestrial application (T-ISDB), it is proposed to adopt a transmission scheme that uses band segmentation with orthogonal frequency division multiplexing (BST-OFDM) [25].

**Digital Radio Mondiale (DRM):** a world-wide consortium of broadcasters and manufacturers. The main system requirements for a digital broadcasting system operating below 30MHz have been defined by the DRM group as:

A system based on a single non-proprietary world-wide standard that offers significant audio quality improvement over analogue short-wave and capable of multimedia operation [25].

**FM In-band/On-channel (IBOC)** IBOC is the approach preferred by broadcasters for introduction of terrestrial DAB services in the United States given the lack of spectrum availability for implementation of a new-band system such as Eureka 147.

IBOC systems are designed to be used in the existing broadcast bands and to be compatible with the analog signals currently in use. However, many countries outside the United States are cautions about IBOC systems for the following reasons:

- IBOC systems are expected to cause some degradation to existing analogue services and/or

- reduce their coverage, although to what extent is still unclear;
- IBOC systems have not yet been examined or demonstrated to independent bodies such as the ITU [25].

**MediaFLO** is a proprietary system developed by Qualcomm to deliver broadcast services to handheld receivers using OFDM. Qualcomm intends to roll-out these services in the 700 MHz spectrum in the United States where it holds a license [26].

**MPEG-2** is the video and audio compression standard used by DVDs and to provide many present-day digital television services [26].

**Orthogonal Frequency Division Multiplexing (OFDM)** is a digital transmission technique which places data on hundreds or thousands of carriers simultaneously [26].